\documentclass[letterpaper,11pt,review]{article}   %% LaTeX 2e (preferred)

\usepackage{aop}

\begin{document}

\title{Spin noise spectroscopy: From proof of the principle to applications }

\author{Valerii S. Zapasskii }
\address{Saint-Petersburg State University, Physics Department, Spin
Optics Laboratory, Saint Petersburg, 198504 Russia }

\begin{abstract}
More than 30 years ago, the feasibility of detecting magnetic
resonance in the Faraday-rotation noise spectrum of transmitted
light has been demonstrated experimentally. However, practical
applications of this experimental approach have emerged only
recently thanks, in particular, to a number of crucial technical
ements. This method has now become a popular and efficient tool
for studying magnetic resonance and spin dynamics in atomic and
solid-state paramagnets. In this paper, we present a review of
research in the field of spin noise spectroscopy including its
physical basis, its evolution since its first experimental
demonstration, and its recent experimental advances. Main
attention is paid to the specific capabilities of this technique
that render it unique compared to other methods of magnetic and
optical spectroscopy. The paper is primarily intended for the
experimentalists who may wish to use this novel optical technique.
\end{abstract}

\ocis{270.2500, 260.5430}

\tableofcontents
\section{Introduction}
\label{sec:intro}

In the photonics and information science of recent years, much
attention is paid to the spin-based, rather than charge-based,
electronic systems. This area of research, nowadays referred to as
spintronics, treats spin as a carrier of quantum information and
is considered as a highly promising pathway to new information
technologies \cite{Wolf,Zutic,Elzerman}. This tendency has arisen
a great interest to dynamic properties of spin-systems both in
equilibrium and under different kind of perturbations capable of
controlling spin state of the system. The new technique - spin
noise spectroscopy,  developed in the last decade, refers to the
experimental methods intended for studying spin-systems under
conditions of thermal equilibrium.

The term 'spin noise spectroscopy' (SNS) signifies experimental
technique that implies spectral investigations of spontaneous
fluctuations of spin-system magnetization (spin polarization). Until recently, this
spectroscopy was not considered as a
practical instrument of experimental physics. A few experiments
\cite{Aleks:1981,Sleator,Mitsui}, in which magnetic resonance was
observed experimentally in the spin noise spectrum, were primarily
aimed at demonstration of practical feasibility of such an
approach. Of course, the fact that spin-system in a
magnetic field should display an excess noise at the frequency of
magnetic resonance justified by the
fluctuation-dissipation theorem (see, e.g., \cite{Landau})
was beyond question. Still, fluctuations of this kind,
fundamentally small for macroscopic systems, could be either
hardly detectable, or of no practical sense. A certain interest to
this technique was first revealed by researchers dealing with
fundamental problems of quantum nondemolition measurements (see,
e.g., \cite{Kuzmich,Sorensen}).

For the last several years, however, the situation has drastically
changed, mainly, due to remarkable advances in the digital systems
of data acquisition \cite{Romer,Muller:2010a,Crooker:2010}. At
present, the Faraday-rotation-based SNS, which is considered to be
the most efficient approach to spin noise detection, has found its
niche among other methods of the magnetic resonance spectroscopy and
gradually turns into a standard experimental technique with a wide
and, in many respects, unique range of applications.

In this paper, we briefly outline the main stages in development
of this new branch of spectroscopy starting with the first
experimental observation of magnetic resonance in the
Faraday-rotation (FR) noise spectrum to the most recent
achievements that revealed unique potentialities of this technique
both in the RF and optical spectroscopy of paramagnets. We will
consider basic ideas underlying this experimental technique, will
discuss technical problems associated with its implementation,
and will describe specific properties of the
FR-based SNS that determine its rich informative potential. We
will also consider recent technical progress in this method of
research that has radically changed its place in the contemporary
experimental physics. This paper is intended not to review the experimental
measurements on SNS, but rather to review properties of the noise and spin-noise
spectroscopy that make these measurements possible.

\section {Historical background}

The FR-based SNS combines in itself two essentially different
methods of research. On the one hand, it can be considered as a
branch of the light intensity noise (LIN) spectroscopy and, on the
other, as a specific modification of the optical method of
magnetization detection. To make clear the essence of the FR-based
SNS, it makes sense to consider these two basic methods in more detail.

\subsection{The light intensity noise spectroscopy}

The pioneering experimental work on the LIN spectroscopy that has
arisen a great interest was performed by A. Forrester {\it et
al.} in 1955 \cite{Forrester} (Fig. 1). In that work, the cell with Hg
vapor placed into a magnetic field was excited by an electrodless
microwave discharge, and two Zeeman components of the emission
line 546.1 nm were detected photoelectrically by means of a
specially designed photodetector coupled to a microwave cavity.
The detected spectrum of the light intensity noise was found to
contain a peak at the frequency of magnetic splitting of these two
components. The excess noise of the light intensity related to this splitting was less than $10^{-4}$ of its shot-noise level.
The complexity of the experiment was additionally aggravated by
high frequency of the detected signal ($\sim 10^{10}$ Hz), needed
to provide sufficiently high Q-value of the noise peak. Still, the
signal has been detected, in this experiment, with the
signal-to-noise ratio of around 2, for the accumulation time of
250 s.

\begin{figure}
\begin{center}
\includegraphics[width=10cm]{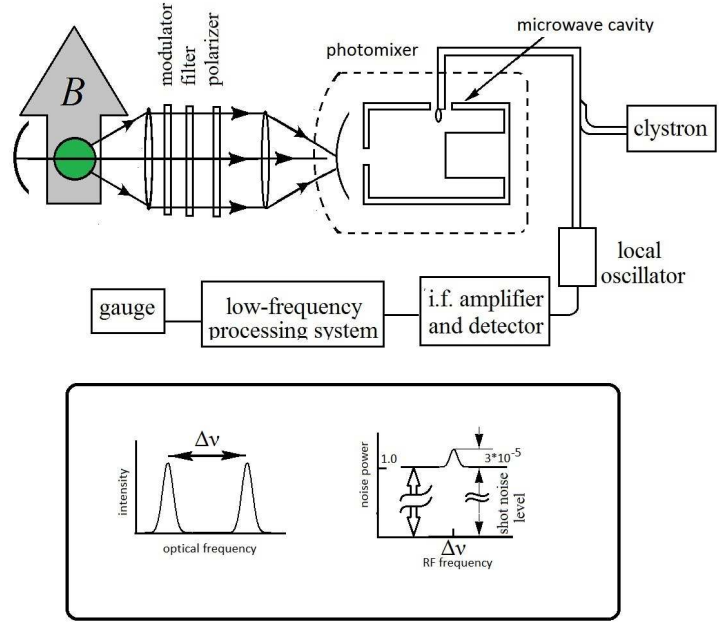}
\caption{Simplified scheme of the experiment of Forrester {\it et al.} \cite{Forrester}. Lower inset shows schematically optical spectrum of the detected Zeeman doublet of Hg vapor (left) and the detected spectrum of the light intensity noise.} \label{Forr}
\end{center}
\end{figure}

It should be emphasized that, in spite of the fact that the frequency
of the observed peak in the intensity noise spectrum exactly corresponded here to
magnetic splitting of the
emission line, this experiment did not have anything to do with optical
detection of magnetic resonance (see \cite{ZapFeof}
and references therein). Suffice it to
say that the width of this peak was determined by the Doppler
broadening of these Zeeman components which should not be revealed in
any way in the magnetic resonance spectrum.

The result of fundamental importance obtained in this paper was
that, in the process of photoelectric conversion of the light
field, the emission probability for electrons is proportional to
the square of the total electric field amplitude arising due to
interference between its Fourier components. In other words, this
paper has shown experimentally the possibility of interference
between the light fields originating from different incoherent
sources. This conclusion, at that time, was far from trivial, and,
as the authors of \cite{Forrester} wrote, ``many physicists found
it contradictory to their ideas about the nature of
interference''.

The discussion raised by publication \cite{Forrester} was further
heated by another important experiment in the field of noise spectroscopy.  The experiment in point was performed and interpreted by Hunbury-Brown and Twiss in 1956 \cite{Hunbury-Brown}. Now, the measurements were focused on detecting spatial correlations in the intensity noise of a remote thermal source and were aimed at application of this effect to evaluating angular dimensions of stars. The authors took advantage of the fact that emission of a remote thermal source of finite sixe is characterized by a certain spatial scale of the light coherence controlled by angular size of the source. Indeed, if we consider intensity relief created by such a source with the diameter $D$ located at a distance $L$ from the plane of observation (Fig. 2), we will see that it comprises a multitude of bright and dark spots. The brightness created at each particular point depends on whether the resultant interference of the waves from all emitters of the source is constructive or destructive. These spots are very pronounced in monochromatic laser fields and are called {\it speckles}. The size of a speckle is determined by the distance at which the total phase difference between the rays providing constructive or destructive interference changes by $\pi$ radians.    In our case, the average size of the spot is given by $L\lambda/D$, where $\lambda$ is the light wavelength. Thus, we see that the angular size of the light source in this geometry can be directly estimated from the average size of speckles.

\begin{figure}
\begin{center}
\includegraphics[width=8cm]{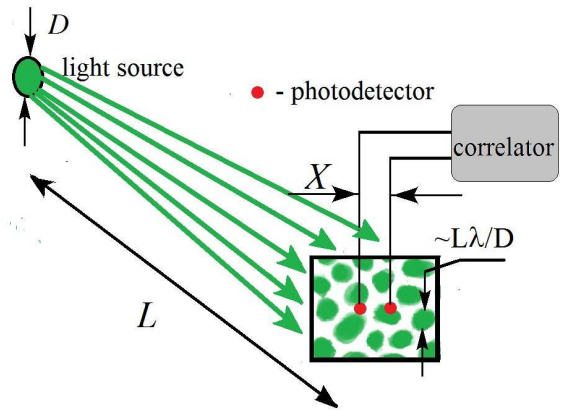}
\caption{Geometry of the experiment of Hunbury-Brown and Twiss} \label{setups}
\end{center}
\end{figure}

Hunbury-Brown and Twiss performed their experiment with a star as a light source and evidently could not examine instantaneous relief of brightness over a large area of the illuminated surface. So, the authors made use of the intrinsic intensity noise of the light emitted by the incoherent source. These temporal fluctuations had to be correlated or not, in two spatial points of the observation plane, depending on whether these points are closer to each other than the size of the spot or not. Thus, by measuring correlation between photocurrents of two detectors placed at a distance of $X$ from each other as a function of $X$ one can estimate the average size of the speckles (of the {\it coherence area}) and to find angular size of the source.

The stellar interferometer created on the basis of this effect (the {\it intensity interferometer}) was successfully used in astrophysical measurements.

The experiments \cite{Forrester} and \cite{Hunbury-Brown} were the earliest and the most famous that have shown that noise in optics may be useful and informative. Note that these measurements were aimed at studying  properties of the light, rather than properties of a medium. At the same time, in 60s-70s, there have been performed a considerable number of experiments in which the intensity-noise-based technique (yet without lasers) was used for studying dynamic properties of atomic systems. Highly important results of these studies were related to possibility of manifestation of the excited state dynamics and fine energy structure in the noise spectrum of spontaneous emission of the system. The noise of the light transmitted by an absorbing atomic medium was employed for measuring diffusion parameters of atoms in gaseous medium. It should be noted that all these experiments, in the pre-laser era, were highly labor-consuming. In more detail, this story is considered in the review \cite{Aleks:1983}.

The situation has changed dramatically, however, with the advent
of lasers. Due to extremely high luminosity of laser emission, the
spectral power density of the light intensity modulation in these
sources at the beat frequencies could exceed the background
shot-noise level by many orders of magnitude, and the LIN
spectroscopy has gained a much deeper sense. It became clear that
the LIN spectroscopy could be used not only for studying spectral
and correlation characteristics of the light source proper (like,
e.g., the laser output mode structure), but also for investigating
dynamic properties of the media interacting with the light.
Specifically, the LIN spectroscopy, referred to as Dynamic Light
Scattering (DLS) or Photon Correlation Spectroscopy (PCS), has
found application for studying morphological and dynamic
properties of suspensions, solutions of macromolecules and
polymers, liquid crystals, biological solutions and
microorganisms. This technique may be efficiently applied to
ensembles of particles with a wide range of dimensions (from 0.001
$\mu$m to several $\mu$m), inaccessible for other methods (see,
e.g., \cite{Berne}).

\subsection{Spectroscopy of the light intensity versus spectroscopy of the light field}

It is interesting to compare the spectroscopy of LIN with the
conventional optical spectroscopy (which also inevitably deals
with fluctuating optical fields). Both methods are aimed at
getting information about spectral characteristics of the optical
field $E(t)$ oscillating at about $10^{15}$~Hz by measuring its
{\it intensity} $|E(t)|^2$ (in reality, by measuring photocurrent
or photocharge), since we cannot directly measure the field
amplitude $E(t)$.

\begin{figure}
\begin{center}
\includegraphics[width=8cm]{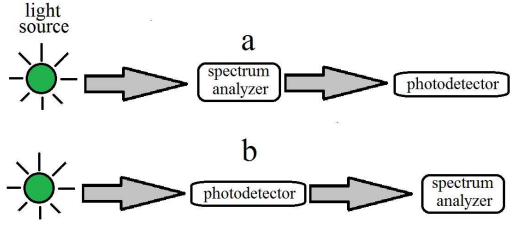}
\caption{The two experimental arrangements used in the conventional optical spectroscopy (a) and in the spectroscopy of light intensity noise (b)} \label{setups}
\end{center}
\end{figure}

 In the standard optical spectroscopy, which implies measuring
the spectrum of the field at optical frequencies and, therefore,
may be considered as the spectroscopy of optical field, the
procedure of spectral decomposition of the light field precedes
measurement of the light intensity. As a result, the intensity of
each detected spectral component retains information about
correlation properties of the optical field which is connected
with its spectrum, according to the Wiener-Khinchin theorem,
through the relationship

\begin{equation}
I_\omega = \int {\langle E(t) E^{*}(t+\tau )\rangle
e^{i\omega\tau} d\tau}
\end{equation}

In other words, optical spectroscopy allows one, in this way, to
get information about correlation properties of the field in the
range of $10^{-15}$ s without photodetectors with so high
temporal resolution.

In the LIN spectroscopy, the experimental arrangement is
inverted: the light field is first converted into a photocurrent
(thus losing its carrier frequency) and then spectrally analyzed
in the electronic channel of the detection system. In this case,
the power spectrum of the signal under study is determined by
correlation characteristics of the {\it light intensity} $I(t)$ (rather than light field $E(t)$)

\begin{equation}
 I_\omega^2 = \int {\langle I(t) I(t+\tau)\rangle e^{i\omega\tau}
d\tau}
\end{equation}

and does not contain, in an explicit form, information about
correlation properties of the field at optical frequencies. In the
LIN spectroscopy, the scale of the frequencies of interest is
limited by the bandwidth of the detection system, and the LIN
spectrum, therefore, can contain information only about relatively
fine features of the optical spectrum.

This fundamental distinction between the two spectroscopic techniques
is schematically illustrated by Fig. 3. The SNS, as a sort of LIN
spectroscopy, evidently implies the second type of the measurements,
with the photodetection preceding spectral decomposition of the signal.
At the same time, combination of the two types of spectroscopy, with
preliminary spectral decomposition of the light field, as we will see
below, may provide valuable additional information about the system.

In some cases, the {\it light intensity spectrum} may provide
information about the light field inaccessible for conventional
optical spectroscopy. The matter is that the light intensity
variations, as was already mentioned, are caused, in fact, by the
beats between different spectral components of the optical field
and, therefore, are able to reveal the phase correlation between
the components not seen or not resolved in the optical spectrum.
The simplest example  may be given by a broadband (or ``white'')
light with harmonically modulated intensity. This light reveals,
in its intensity spectrum, a sharp peak at the modulation
frequency, while its optical spectrum remains practically
unperturbed by the modulation. This informative capability of the
light-intensity spectroscopy constitutes the basis for
spectroscopy of superhigh-resolution. This spectroscopy, developed
in the 70s, made it possible, in particular, to realize
sub-Doppler resolution in optical spectroscopy of atomic systems
and to study fundamental phenomena of interference of quantum
states hidden in the inhomogeneously broadened optical transitions
(\cite {Aleks:1972,Aleks:1979}).

 The FR-based spin noise spectroscopy (SNS), we will talk about,
can be regarded as a polarization version of the LIN spectroscopy,
with the light field fluctuations provided by spontaneous noise of
the spin-system magnetization.  Possibility of conversion of the
magnetization fluctuations to those of the light polarization is
determined by direct relationship between the Faraday rotation and
magnetization of the spin-system, constituting the basis of the
optical method of magnetic measurements (see, e.g.,
\cite{Zap:1987}).

\subsection{Optical detection of spin-system magnetization}

Optical methods of detecting spin-system magnetization (spin polarization) employ
the fact that magnetization of a paramagnet affects its optical properties.
The so-called ``paramagnetic''  FR \cite{B&S} that directly
reflects magnetization of the spin-system is formed in the
following way (Fig.4). The difference between populations of the
ground-state magnetic sublevels of the paramagnet (of the
spin-system energy levels) creates, for the light propagating
along the field, corresponding difference in the optical
absorption for the two circular polarized transitions from these
unequally populated states ({\it magnetic circular dichroism},
MCD). This dichroism is, evidently, observed only in the region of
optical absorption and manifests itself in the form of ellipticity
acquired by the linearly polarized light transmitted through the
medium. The same kind of optical anisotropy, in conformity with
the Kramers-Kronig relations, is observed in difference of the
refractive indices for the two circular polarizations ({\it magnetic
circular birefringence}) and is revealed as rotation of the
polarization plane of the linearly polarized light traveling
through the medium  ({\it the Faraday effect}).

\begin{figure}
\begin{center}
\includegraphics[width=10cm]{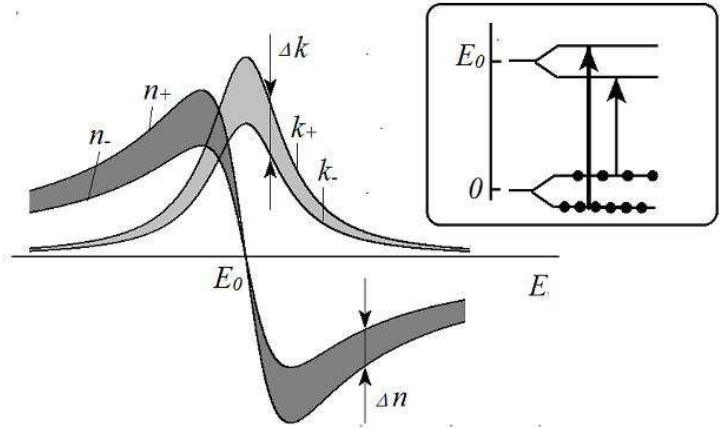}
\caption{ Formation of ``paramagnetic'' part of the Faraday
rotation In a longitudinal magnetic field for the simplest case of
transition between two magnetic doublets (inset). Due to
redistribution of populations  over the ground-state sublevels,
the absorption coefficients ($k_+$ and $k_-$) and refractive
indices ($n_+$ and $n_-$) for two circular polarizations become
different. These differences ($\Delta k$ and $\Delta n$) give rise
to the magnetic circular dichroism and Faraday rotation,
respectively. Magnetic splitting of the transition energies is
supposed negligible compared to the line width. } \label{fig1}
\end{center}
\end{figure}

In this description, we ignored magnetic splitting of the transitions into circularly polarized components of different handedness.  This contribution to the magneto-optical activity of the medium (usually referred to as ``diamagnetic'') is not related to spin-system magnetization and is irrelevant to our consideration.

Thus, magnetization of a paramagnet can be measured by detecting
either circular dichroism or Faraday rotation. An essential
difference between  these two methods is that the method of
dichroism implies inevitable optical excitation of the sample and,
therefore, cannot be nonperturbative, whereas detection of the
Faraday effect may be performed in the region of transparency
without producing any real optical transitions. In addition, the MCD-based measurements
of spin-system magnetization may be performed in a pure photometric way be measuring intensity of the transmitted
circularly polarized light, while the Faraday rotation in the transparency region does not imply ant changes in the light intensity and requires the use of polarimetric technique.

It is also worth to note that  the Faraday rotation, in the context of our treatment, is not necessarily to be observed in the longitudinal magnetic field, as implies its classical definition. It is supposed that the FR detects spin polarization along the light propagation direction regardless of the direction (and
even of the presence) of the external magnetic field. In particular, the FR can be used to detect oscillation of transverse magnetization of the spin-system under conditions of its coherent precession and, thus, to optically detect its magnetic resonance. This method was first proposed and realized in \cite{Bitter}.

\section{ Basic idea of the Faraday-rotation-based spin noise spectroscopy }

Application of magnetooptics to studying spin dynamics is known
since 50s of the last century. The FR and MCD effects were used,
in particular, for detecting magnetic resonance in atomic systems
polarized by optical pumping. In those experiments, the magnetic
resonance was revealed either as a suppression of longitudinal
magnetization (MCD or FR) for the light propagating along the
field) or as oscillations of the transverse magnetization (MCD or
FR for the light propagating across the field) \cite{Dehmelt,BB}.
This magneto-optical technique was also used for studying dynamic
properties and energy structure of transparent paramagnets by
measuring their magnetic susceptibility at subresonant frequencies
\cite{ZKM}.

It is important that, in all these experiments, the detected
optical signal was a result from {\it coherent response} of the
spin-system to a modulated (or, at least, time dependent) perturbation (microwave pumping or
external magnetic field). The idea of the spin noise spectroscopy
was to detect intrinsic fluctuations of the spin-system
magnetization under conditions of thermal equilibrium without any
regular external perturbation.

Since any real spin-system consists of a finite number of spins
which perform incessant random motion (maintaining thermal
equilibrium of the system), its magnetization should inevitably
exhibit spontaneous fluctuations due to deviations of
instantaneous values of the magnetization from its mean value (Fig.5).

\begin{figure}
\begin{center}
\includegraphics[width=8cm]{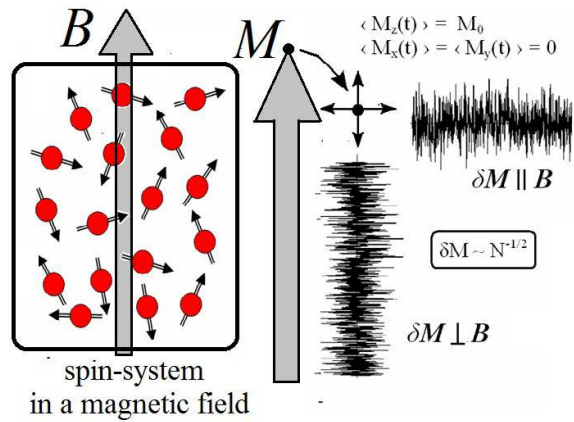}
\caption{Magnetization ${\bf M}$ of a spin-system in a static external magnetic field ${\bf B}$, due to permanent motion of individual apins,  exhibits random fluctuations both in its magnitude (along the magnetizatio ${\bf M}$) and in its  direction (across the mean magnetization).}  \label{fig3}
\end{center}
\end{figure}

Spectrum of these random fluctuations and their correlation times
should be directly connected with dynamic parameters of this
motion (characteristic frequencies and correlation times). These
dynamic characteristics, in principle, can be extracted from
results of the noise measurements.

Of course, the notion of thermodynamic fluctuations had been known
long before the late 70's, when the experiment
 \cite {Aleks:1981} was planned, and the problem was
more of practical, rather than fundamental nature: it was clear
that these fluctuations, in a macroscopic spin-system, should be
extremely small, and the question was whether it was possible to
detect them experimentally for a reasonable accumulation time with
a sufficiently high signal-to-noise ratio and, after all, whether
it had any practical sense.

\section {Experimental geometries}

Conventional magneto-optical measurements imply two main
geometries of the measurements, with the light beam propagating,
respectively, along and across the magnetic field. These
geometries are commonly referred to as the Faraday (longitudinal)
and Voigt (transverse) configurations. Standard magneto-optical
effects in a static magnetic field, observed in these
configurations, in conformity with the symmetry of the problem,
correspond to the cases of magnetic-field-induced circular and
linear anisotropy. This is not the case, however, for fluctuations
of the magneto-optical anisotropy, which may break the symmetry of
the problem.

\begin{figure}
\begin{center}
\includegraphics[width=10cm]{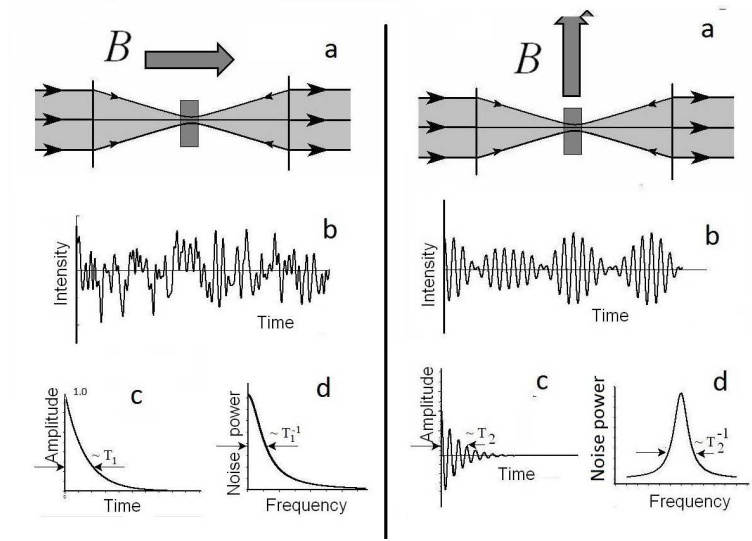}
\caption{Measuring the spin noise (FR noise) of a spin-system in an external magnetic field. The left and right panels correspond to longitudinal and transverse orientation of the magnetic field with respect to the light propagation (the Faraday and Voigt configurations, respectively). (a) - Experimental arrangement, (b) - a sketch of temporal dependence of the noise signal, (c) shape of  the correlation function, and (d) -
spectrum of the noise signal. } \label{fig2}
\end{center}
\end{figure}

\subsection{Faraday configuration}

The simplest experimental geometry for measuring the FR noise,
which may seem the most natural, implies detection of {\it
longitudinal} fluctuations of the magnetization, with the probe
laser beam aligned along the external magnetic field and along the
equilibrium magnetization of the spin-system (Faraday
configuration, Fig. 6, left panel). In this case, the detected fluctuations
of the magnetization do not break axial symmetry of the system.
Dynamics of these fluctuations is controlled by the only relevant
characteristic time - longitudinal spin relaxation time $T_1$,
which determines the correlation and spectral characteristics of
this random process. In the time domain, the longitudinal FR noise
(Fig. 6,b, left panel) looks as a ``white'' noise with the removed higher
frequencies (exceeding the relaxation rate $T_1^{-1}$). The
autocorrelation function, correspondingly, acquires exponential shape
(rather than $\delta$-wise, as for the ``white'' noise) with the
characteristic time $T_1$ (Fig. 6,c, left panel), while the power spectrum of
the process, given by Fourier transform of the autocorrelation
function, is a Lorentzian centered at zero frequency with the
width determined by the relaxation rate $T_1^{-1}$ (Fig. 6,d, left panel).

Thus, by measuring the FR noise spectrum in the Faraday geometry,
we can obtain information about time $T_1$, its dependence on
magnetic field, temperature, and other external parameters, i.e.,
about properties of the system usually obtained from the data of EPR
spectroscopy or from nonresonant magnetic measurements in
oscillating magnetic fields.

It is worth to note that this experimental approach has much in
common with the paramagnetic relaxation technique which has been
developed by the Dutch physicist C. Gorter \cite{Gorter} for
measuring magnetic susceptibility of paramagnets in parallel
fields and was later transferred to optical basis with the aid of
laser polarimetry \cite{ZKM,Aleks:1978}. In Gorter's technique,
dynamics of the spin-system is studied by measuring its linear
response to a RF magnetic field applied along the dc field. The
essential difference between these two approaches is that in the
case of noise spectroscopy, the system is not supposed to be
perturbed by the external oscillating field. In terms of the
fluctuation-dissipation theorem, the longitudinal magnetization
noise (or the FR moise) is the counterpart of the linear magnetic (or magneto-optic)response observed in
parallel fields.

In this longitudinal geometry, however, the noise spectrum is located in the vicinity of zero frequencies
where the measurements are often hampered by the universal $1/f$
(flicker) noise. In addition, spectrum of the longitudinal
fluctuations of magnetization, bringing much data about relaxation characteristics of the system, does
not contain any information about its magnetic splitting, which is highly important for characterization of the spin-system.

\subsection{Voigt configuration}

The above drawbacks of the Faraday configuration can be easily overcome by passing to the Voigt
geometry (Fig. 6, right panel). In this cobfifuration, the light beam traveling through
the paramagnet across the applied magnetic field will detect
transverse magnetization of the spin-system, which, by symmetry
considerations, should vanish in this geometry and may arise only due to spontaneous violation of  the symmetry by fluctuations.

In this configuration, any random transverse fluctuation of
magnetization will precess around the magnetic field direction
during the transverse relaxation time $T_2$ (or what is usually
called $T_2^*$, when the spin-system is inhomogeneous) and then
will be replaced by another realization of the transverse
fluctuating magnetization with another magnitude and another phase
of precession. As a result, the probe light beam will exhibit
randomly oscillating FR at Larmor frequency (Fig.6,b, right panel). Correlation
function of this random process will now have the shape of
oscillatory decay with the characteristic time $\sim T_2$ (Fig.
6,c, right panel). The  peak of the FR noise spectrum will now be shifted away
from zero frequency, and the width of the peak will be determined
by the dephasing time $T_2$ (or $T_2^*$) of the spin-system (Fig.
6,d, right panel). In other words, the FR noise spectrum will present, in this
geometry, the {\it magnetic resonance spectrum} of the
spin-system.

The idea of detecting magnetic resonance in the noise of the
spin-system magnetization was first mentioned by F. Bloch
\cite{Bloch} in 1946. An essential contribution to understanding
of optical manifestations of spin-system magnetization, spin
precession, and spin dynamics and, implicitly, to optical
detection of magnetization noise, was made by A. Kastler in his
works on optical pumping \cite{Kastler1,Kastler2,Happer}. The
FR noise spectroscopy is closely related
to the time resolved Faraday rotation \cite{Crooker-97} and may be
considered as its incoherent version.

\section {On polarimetric sensitivity}

A specific feature of the SNS is that it is, in fact, not a {\it
spectroscopy of response} \footnote{To a certain extent, it may be
considered as a spectroscopy of response of a spin-system to its
stochastic perturbation by thermal reservoir of the environment},
and, therefore, the magnitude of the detected signal ({\it noise
signal}) cannot be controlled by varying the strength of the
perturbation. In addition, magnitude of the magnetization (and FR)
noise, for a macroscopic sample, as was already mentioned, should
be extremely small as compared with the values of ``coherent''
magnetization (or magnetization of saturation) induced, by the external perturbation acting in the same way upon all the spins of the ensemble. Thus, it is clear
that the problem of polarimetric sensitivity may have a critical
importance for the spin noise spectroscopy.

The sensitivity of polarimetric measurements in optics is known to
be fundamentally limited by the so-called photon noise or shot noise of the detector photocurrent in accordance with the relationship (see, e.g.,
\cite{Zap:1982})
\begin{equation}
 \Delta\varphi_{min} \approx \sqrt{\Delta{f}/{2I\eta}}
\end{equation}
($\Delta\varphi_{min}$ is the angle of the polarization plane
rotation detected with the signal-to-noise ratio equal to unity,
$\Delta{f}$ is the bandwidth of the detection channel, $I$ is the
light intensity measured in the number of photons per second
\footnote{Everywhere below, when the issues of polarimetric
sensitivity are discussed, the light intensity $I$ is supposed to
be given in photons per second.}, and $\eta$ is the quantum yield of
the photodetector). For instance, for the light power 20 mW,
wavelength 550 nm, and bandwidth 1 Hz, this quantity lies in the
range of $10^{-8}$ rad.

Such a sensitivity, as may be shown by appropriate estimates, is high enough to solve, in many
cases, the problem of spin noise detection. The most frequent
reason why the shot-noise-limited sensitivity cannot be achieved
with the lasers (which are considered, in this context, the only
suitable light sources) is related to their excess noise, which
may exceed the shot-noise level by a few orders of magnitude.

The excess noise can be suppressed using different expedients
\cite{Zap:1982}, the simplest and most efficient of them being
{\it balanced detector} (Fig.7,a).

\begin{figure}
\begin{center}
\includegraphics[width=12cm]{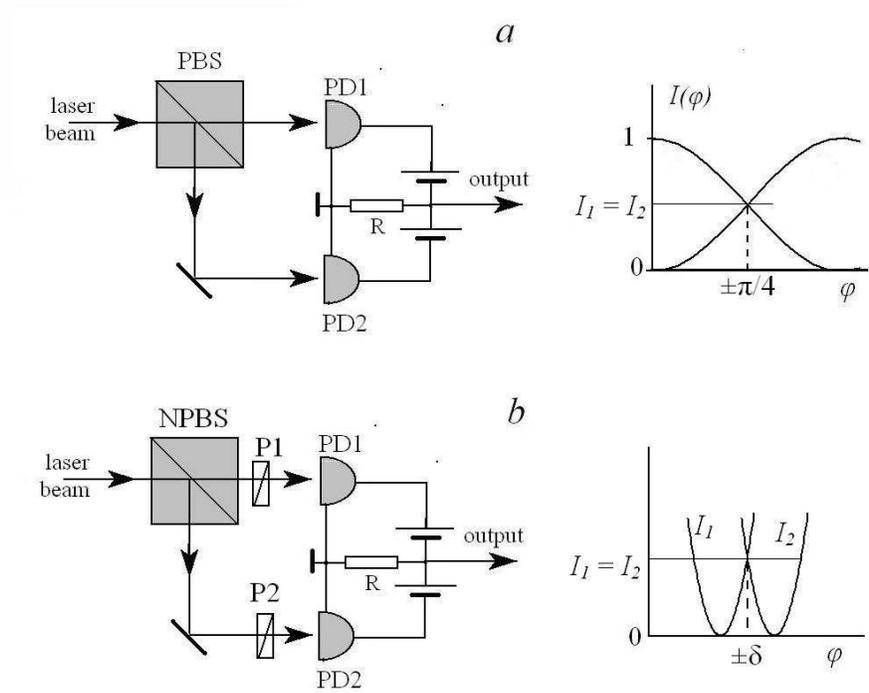}
\caption{Two schemes for measuring polarimetric signals with a
balanced detector (left side) and behavior of the detector photocurrents versus azimuth of the polarization plane of the
incident light $\varphi$ (right side). (a) -- Standard $45^0$-geometry. At $\varphi = 45^0$, the
photocurrents become equal, and the current flowing through the
resistor $R$ vanishes for any light intensity. (b)
 -- The scheme with variable polarization extinction. The beam is
split by a non-polarizing beamsplitter (NPBS), retaining polarization of the light beam, and the level of
polarization extinction is set in each arm independently (using polarizers $P1$ and $P2$) to
equalize photocurrents. At high extinction (at small angles
of detuning $\delta$ from the crossed position), the steepness of
the dependence $I(\varphi)$ may become much higher.  PD1 and PD2 -
photodetectors, P1 and P2 - polarizers. } \label{fig4}
\end{center}
\end{figure}

The linearly polarized light beam whose polarization behavior is
examined passes through a polarization beamsplitter with its two
outputs coupled to two photodetectors. The photodetectors are
included into a differential circuit so that their photocurrents
are subtracted at the load resistor $R$. When the polarization plane
of the incident light makes an angle of $45^0$ with polarizing
directions of the beamsplitter, photocurrents of the two detectors
cancel at the resistor (regardless of the light intensity and
its variations).  At the same time, the changes of the
photocurrent related to variations of the light beam polarization
plane are always anticorrelated (have opposite signs) and, as a
result, are summed up. In this way, it is possible, in practice,
to suppress the excess intensity noise by approximately three
orders of magnitude and to achieve the shot-noise limit of
polarimetric sensitivity with noisy laser sources.

For the first time, as far as we know, the shot-noise-limited
polarimetric sensitivity has been realized (with the use of the
balanced detector), in \cite{Aleks:1976,Jones}. Nowadays, the
balanced detectors are produced commercially and, in combination
with polarization beamsplitters, are commonly used in the
high-sensitive polarimeters.

There are some other interesting methods of the excess noise
suppression based on the fact that the shot-noise-limited
polarimetric sensitivity does not substantially change when the
angle between the analyzer and polarization plane varies from
$\pi/4$ to total extinction. Indeed, when moving towards the total
extinction (let it be $\varphi = 0$), the polarimetric response
($\Delta I$) varies approximately linearly with $\varphi$ ($\Delta
I \sim I_0\sin\varphi \cos\varphi\Delta\varphi$). At the same
time, the transmitted light intensity at small $\varphi$ varies as
$\varphi^2$  ($I \sim I_0\sin^2\varphi$) providing the shot noise
(varying  as $\sqrt{I}$ ) also proportional to $\varphi$. Thus,
the signal-to-noise ratio remains in this range of angles
$\varphi$ practically the same (to within the factor $\sin{\pi/4}
= 1/\sqrt{2}$). In reality, however, this conclusion is violated
at small $\varphi$ either because of nonideality of the
polarization system (the extinction ratio of the polarizers
usually cannot be smaller than 10$^{-5}$ -- 10$^{-6}$, or  when
the detected light intensity noise sinks in the noise of
electronics.

In the absence of excess fluctuations, capable of reducing
polarimetric sensitivity, the high-extinction polarization
geometries can still be used for other purpose, namely, to raise
the probe beam intensity (and thus to increase the
shot-noise-limited polarimetric sensitivity) leaving the input
light power on the detectors at low level. This is especially
important for the SNS, which employs broadband photodetectors with
small photosensitive area unable to endure high input power.

One of the methods that allows one to realize the high-extinction
polarization geometry and, at the same time, to retain opportunity
to further suppress the excess intensity noise is based on the use
of polarization pile, which serves, in this case, as a dichroic
medium with no birefringence. As has been shown in
\cite{Zap-pile}, the polarization pile makes it possible to
considerably magnify the polarization plane rotation angle at the
expense of reduction of the beam intensity. This gadget can also
be used for choosing favorable ratio between intensity of the
light passing through the sample and the input light power of the
photodetector with no loss in sensitivity.

Among other high-extinction polarization geometries can be
mentioned the scheme with polarization-insensitive beamsplitter
and independent adjustments of polarization extinction in two
channels (Fig.7,b). This method has all the merits of the
polarization pile, but can be easier realized experimentally.

As applied to measurements of the spin-system magnetization noise,
there are certain additional requirements that should be met. In particular, the wavelength of the probe laser
beam should provide the most efficient conversion of the
magnetization noise to that of the FR (this wavelength does not
necessarily coincide with the region of the greatest Verdet
constant of the system, see below). The other requirement is
related to maximizing FR per a single spin, which can be achieved
by reducing cross section of the probed volume and increasing its
length.

Indeed, let we probe a paramagnet with the length $l$ and spin
density $n_0$ by the light beam with the cross section $S$ (all
quantum systems of the ensemble are supposed to be identical).
Then, the number of spins confined in the beam will be $n =
n_0lS$. Fluctuations of the Faraday rotation at the exit of the
paramagnet will be proportional to fluctuations of this quantity
($n^{-1/2}$) and to the total, independent of $S$, Faraday rotation ($\sim n_0l$). As a result, for
the FR noise, we have
\begin{equation}
 \Delta\varphi \sim n_0l/\sqrt{n_0lS} = \sqrt{n_0l}/\sqrt{S}
\end{equation}
So, for a given paramagnetic sample, the FR noise varies in
inverse proportion with the radius of the light spot
\cite{Crooker:2004,Muller:2010b}. This is a highly important
feature of the SNS, which implies that the spin noise (with all
other factors being the same) {\it increases} with decreasing
number of spins. This is why, in the SNS, the probe laser beam
passing through the sample should be preferably tightly focused.

Note that, using the light beam for detecting spin fluctuations, we
probe a fluctuating medium (spin-system) by a fluctuating agent
(photon flux) and measure intensity fluctuations of the light on
the background of its own shot noise \cite{Sorensen}. It may seem, at
first sight, that the spin noise can be detected only when the
noise introduced by the spin-system exceeds (or is comparable
with) the shot noise of the probe beam. In reality, however, this is not the
case. As can be shown, the mean-square level of Poissonian noise
transmitted through a Lorentzian filter with the bandwidth
$\gamma$ for the accumulation time $\tau$ can be measured with a
relative accuracy of $[(1 + \gamma e/i_{ph})/ \gamma\tau]^{1/2}$
($i_{ph}$ is the photocurrent and $e$ is the electron charge),
which turns into the known factor $(\gamma\tau)^{-1/2}$ at
sufficiently large photocurrents ($I_{ph} \gg \gamma e$)
\cite{Kharkevich}. So, one can easily estimate that, under real
experimental conditions and for sufficiently large accumulation
times, the factor $(\gamma\tau)^{1/2}$ may reach many orders of
magnitude. As a result, the shot-noise power appears to be defined
with a fairly high accuracy and, in the form of a stable spectral
background, does not preclude measuring intensity fluctuations
whose power lies essentially below the shot noise level.

\section{The first experiment}
\subsection{Magnetic resonance in the FR noise spectrum }

The first experiment on detection of magnetic resonance in the FR
noise spectrum \cite{Aleks:1981} was performed on sodium atoms in
 the atmosphere of buffer gas (neon, $\sim$ 10 Torr). Schematic
of the experimental setup is shown in Fig.8.
\begin{figure}
\begin{center}
\includegraphics[width=12cm]{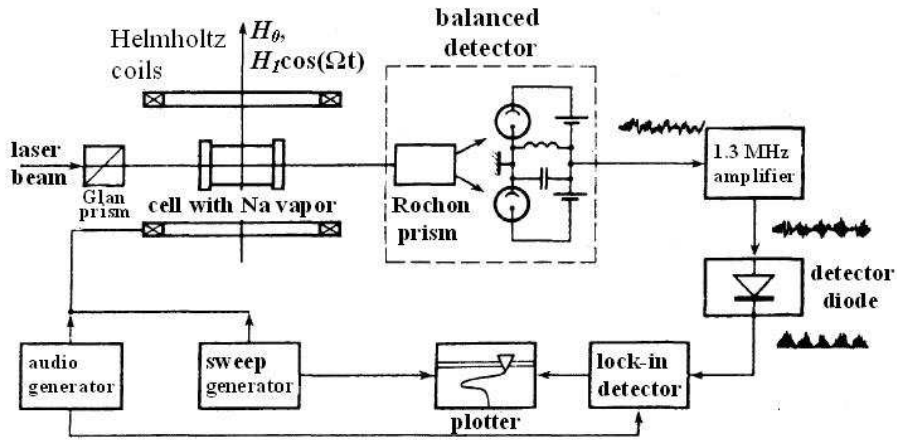}
\caption{ Schematic of the experimental setup used for detecting
magnetic resonance in the Faraday rotation noise spectrum of
sodium atoms \cite{Aleks:1981}. } \label{fig5}
\end{center}
\end{figure}
As a light source, we used a cw dye laser tuned to close vicinity
of the D1 or D2 absorption line of Na. The cell with sodium vapor
was placed into a transverse magnetic field created by a pair of
Helmholtz coils. The field was slightly modulated at a frequency
of $\sim 10^2$ Hz. The differential signal of the balanced
photodetector was first filtered by a resonant circuit at a
frequency of 1.3 MHz, then selectively amplified, quadratically
detected (rectified), lock-in amplified at the field modulation
frequency, and recorded as a function of the applied magnetic
field. In other words, the magnetic resonance frequency was swept
with respect to resonance frequency of the photodetection system
(much like the resonance frequency is swept with respect to
frequency of the RF or microwave oscillator in conventional EPR
spectroscopy).

\begin{figure}
\begin{center}
\includegraphics[width=8cm]{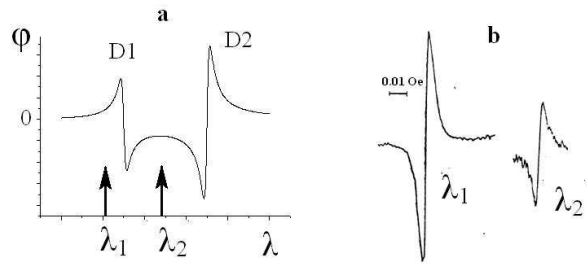}
\caption{ A sketch of the Faraday rotation spectrum of sodium
atoms (a) and experimental plots of the EPR signal in the Faraday
rotation noise (b) for two wavelengths of the probe beam
(indicated by arrows). Due to modulation of the applied magnetic field, the signal, in this experiment, was
proportional to derivative of the FR noise power with respect to
magnetic field. } \label{fig6}
\end{center}
\end{figure}

 Figure 9 shows
two typical plots of the detected spin-noise resonance. The probe
beam was tuned to the edge of D1 line or to the midpoint between
the two lines. Polarimetric signal of the FR noise, in this
experiment, substantially exceeded the shot noise level, and the
signal-to-noise ratio (or, better to say, the ratio of useful
noise to harmful noise) for the accumulation time 2 s, was close
to 100:1. This was the first experimental demonstration of
feasibility of the spin noise spectroscopy. \footnote{Our attempt
to submit the manuscript of \cite{Aleks:1981} to {\it Optics
Communications} under the title "A New Optical Method for ESR
Detection" had failed. The manuscript was rejected on the grounds
that advantages of the proposed technique over conventional
methods of the ESR detection seemed dubious.}

\subsection{Nuclear spin noise}

In 1985, there has been performed the experiment \cite{Sleator} on
nuclear spin noise, which did not have anything to do with the FR
technique, but ideologically was identical to the experiment
\cite{Aleks:1981}. As a spin-system, in that experiment, was used
an ensemble of nuclei $^{35}$Cl in the NaClO$_3$ crystal. The
experiment was performed at 4 K, and the magnetization noise was
detected in a straightforward way using a SQUID sensor. This work
has considerably contributed to the SNS as a first observation of
the resonant spin noise in a nuclear spin-system predicted by F.
Bloch \cite{Bloch}. Here, the number of spins contributed to the
signal was much greater than in the experiment with electron spins
\cite{Aleks:1981}, and the noise signal was expected to be
extremely small. Indeed, the peak of spin noise at the frequency
of nuclear quadrupole resonance of $^{35}$Cl was detected with the
signal-to-noise ratio of a few units for the accumulation time of
7 hrs. Still, further technical advances in this area made it
possible to observe the nuclear spin noise at room temperature
\cite{McCoy,Gueron} and even to develop, on its basis, an
efficient method of nuclear spin-noise imaging \cite{Muller:2006}.
This technique, though less sensitive than the conventional method
of NMR imaging, may be indispensable, in certain cases, as an
entirely noninvasive tomography that does not use any external RF
or X-ray fields.

\section{Evolution of the spin noise spectroscopy }

The effect of magnetic resonance in the FR noise spectrum remained
practically unnoticed, as a possible experimental tool of
magnetic-resonance spectroscopy, until the verge of our century,
when, with some modifications, it was reproduced on other alkaline
atoms and then applied to semiconductor systems. Below, we will briefly outline
the main steps in the development of the experimental
SNS. .

\subsection {Application to atomic systems }

In 2000, an interesting experiment was performed by T. Mitsui
\cite{Mitsui}, who independently observed optically detected spontaneous spin noise
of rubidium atoms at the Larmor frequency.  The
idea of the experiment was the same, but realized differently: the
magnetization noise was detected in the optical absorption, rather
than in the Faraday rotation, of the atomic vapor.

A circularly polarized beam of a diode laser, tuned in resonance
with the D1 absorption linea of $^{85}$Rb (5 $^2$S$_{1/2}$ -
$^2$P$_{1/2}$) was transmitted through the cell with atomic vapor
at 80$^0$C placed into a transverse magnetic field. To simplify
qualitative analysis of the results, the transition was strongly
broadened (up to 6 GHz) by a buffer gas (nitrogen) at 200 Torr.
Fluctuations of the spin-system magnetization, in this experiment,
were observed in the intensity (rather than FR) noise spectrum of
the transmitted light.  A specific feature of this approach was
that the excess intensity noise of the light source, in this case,
should have been eliminated before the sample, rather than
suppressed after it, because, otherwise, it could induce nonlinear
effects undesirable in this experiment (they were studied
independently). The laser source used for
these measurements was characterized by extremely low excess
intensity fluctuations, and, therefore, there was no need to employ balanced detector for their suppression.

In that experiment, fluctuations of the spin-system magnetization
were, in fact, detected by the noise of the MCD (rather than FR)
of the paramagnet. From the viewpoint of linear optics, these two
quantities (MCD and FR), being connected through the
Kramers-Kronig relations, are identical and should give the same
results unless the optical excitation affects spin dynamics of the
system. At the same time, in the MCD-based measurements, each act
of absorption interrupts spin precession of the atom, and it may
seem that the later absorption events cannot be correlated, in any
way, with the previous. The author has shown, however, that this
is not the case for a thermodynamically fluctuating spin-system,
and the MCD-based method of the spin noise detection, which looks
more perturbative than the one based on the FR,  is also
applicable to macroscopic systems.

It is worth to note here that the question about
possibility of detecting superposition states of a quantum system
(as, e.g., a precessing spin) in the noise spectrum of electromagnetic field is not trivial.
In particular, it has been shown in \cite{AKK,AK} that the spontaneous
emission noise spectrum of an atomic medium excited randomly, in a
Poissonian way, does not reveal any features related to its decay kinetics.
It means that the intensity fluctuations of spontaneous emission excited
stationary in this way cannot be used for detecting spin noise in the excited
state of the atoms. A rigorous quantum-mechanical description of the intensity-fluctuation spectrum of spontaneous emission is given in \cite{Smirnov}.

In 2004, S. Crooker et al. \cite{Crooker:2004} examined in more
detail the spin-noise spectra of Rb and $^{41}$K atoms using the
FR-based method and provided a deeper insight into capabilities of
this technique. It was shown, in particular, that magnetic
resonance in the FR noise spectrum can be also observed in the
Faraday configuration (in longitudinal magnetic field). This is
possible, in conformity with the fluctuation-dissipation theorem,
for the ESR transitions allowed in the ac magnetic field
oscillating along the external dc field. An illustration of such a
situation was presented in \cite{Crooker:2004,Mihaila}, where the
resonances of this type were originated from transitions with
$\Delta m_F = 0 $ ($m_F$ is the projection of the total angular
momentum of the state) in $^{41}$K atoms.

\subsection{Perturbative or not?}

One of the questions that arose after successful detection of spin
noise in atomic gases, with strongly off-resonant probing of the
system, was whether this technique could be considered as
nonperturbative or not. At first glance, it looked fairly passive
or nonperturbative because it utilized Faraday rotation in the
transparency region, the probe beam did not induce any real
electronic transitions and, therefore, did not disturb the
spin-system under study. On the other hand, however, it could not
be nonperturbative for evident reasons: the initially
monochromatic probe light, after passing through the medium,
became modulated and thus acquired sidebands with shifted
frequencies. It means that some photons of the beam experienced
inelastic interaction with the medium inevitably accompanied by
some energy exchange between the light and the spin-system. It is
also known that the effect of transparent medium on polarization
of the light beam is accompanied by the back action of the light
upon the medium even in static magnetic field \cite{Happer2}.

This contradiction was resolved by Gorbovitskii and Perel
\cite{Perel} who have shown that the noise of the FR at the
frequency of Larmor precession can be considered as a result of
{\it coherent forward Raman scattering} of the probe laser light
by the ensemble of spins randomly precessing in the transverse
magnetic field. The oscillating polarization in the probe beam arises dues to interference of the forward-scattered light with the light directly transmitted by the medium. This effect, according to \cite{Perel}, is analogous to the effect of optical mixing used in \cite{Eden} for studying Brilloin scattering.

So, this process can be regarded as nonperturbative, because the
probe light interacts with macroscopic fluctuations of the system,
and does not select any particular spin to flip. Such a
conclusion is supported by the possibility to observe magnetic
resonance in the MCD noise, when each absorption event explicitly
destroys coherent superposition of the ground-state sublevels. The
technique evidently becomes more and more perturbative as we pass
from macroscopic ensembles to microscopic ones.

Note that in semiconductors, which are more complicated than pure
atoms and cannot be treated in the framework of so simple models,
a small residual absorption may dramatically affect properties of
the spin-system under study even in the regions of nominal optical
transparency \cite{Crooker:2009}.

The atomic systems, mainly alkali atoms, as the most convenient
model objects of SNS, have been  widely used in fundamental
research related to non-demolition measurements and interaction
between squeezed states of light and matter. It was shown, in
particular, that the spin-noise measurements performed in the
region of optical transparency make it possible to surpass the
standard quantum limit of phase measurements and thus to produce
the squeezed atomic spin states (see, e.g.,
\cite{Kuzmich:98,Kuzmich:00}). This interesting topic and
associated experimental results, however, lie outside the scope of
this review.

\subsection{``Active'' spin noise spectroscopy}

It is worth to mention here another experimental approach to the
optical detection of incoherent spin precession, usually also
regarded as a sort of noise spectroscopy
\cite{Mitsui,Yabuzaki,McIntyre,Ito,Martinelly,Valente}. This
approach is based on transformation of modulation spectrum of the
light transmitted through a paramagnet in the Voigt geometry. In
this case, the light beam serves simultaneously as a probe and as
a pump, and its modulation (either in intensity or in
polarization) is supposed to be equivalent to modulation of the
effective magnetic field applied along the light propagation. This can be achieved either by intensity
modulation of a circularly polarized beam or by modulating the
degree of its circular polarization. Experimental setup for these
measurements may look as shown in Fig. 10.

\begin{figure}
\begin{center}
\includegraphics[width=8cm]{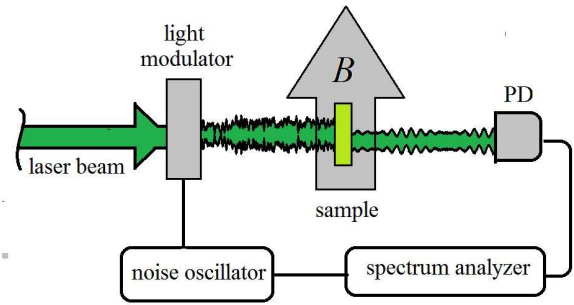}
\caption{ Schematic of the experimental setup for the ``active''
spin noise spectroscopy. In contrast to the conventional SNS,
the transmitted light is not subject to polarization analysis, the beam is
not necessarily focused on the sample, and no intensity noise
suppression is used. } \label{fig7}
\end{center}
\end{figure}

It is well known that the coherent Larmor precession of a
spin-system can be induced optically by the resonant pump light
modulated at the appropriate frequency. The first experimental
observations of the optically driven spin precession were
announced in \cite{BB1}. The effect was detected in the vapors of
alkali metals (Cs and Rb) and in metastable helium. The cell was resonantly pumped by circularly polarized light propagating
across the applied magnetic field and modulated
in intensity at the frequency close to that of Larmor precession
of the spin-system. The resonance was observed as a peak in the
transmitted light intensity at the point where the modulation
frequency coincided with that of the ground-state Zeeman splitting.  This effect of
optically driven spin precession, often referred to as `optical
orientation in the rotating coordinate frame' or as a `resonance of
coherence', is very close, in its physical content, to the effects
of coherent population trapping \cite{Alzetta} and
electromagnetically induced transparency \cite{Harris}. All these
effects imply excitation of coherent superposition of two states
(in our case - of two low-lying states) by superposition of two
optical fields (or by a single modulated field, which is
practically the same).

The ``active'' spin noise spectroscopy exploits the same ideal,
but uses, for this purpose, the pump modulated by more or less
``white'' noise (rather than harmonically). Under these
conditions, the system is offered, so to say, to choose by its own
the frequency component capable of inducing its Larmor precession.
This approach utilizes the effects of nonlinear optics and is
evidently essentially perturbative.

Regularities of the ``active'' SNS strongly differ from those of
the FR-based spectroscopy of spontaneous spin noise. At the same
time, in certain respects, it can be more convenient in practice
and may provide additional information about the system not
related directly to its magnetic properties.

From the viewpoint of experimentalist, this type of noise
spectroscopy is, in many respects, the exact opposite of the
FR-based SNS: it does not imply high polarimetric sensitivity and
the laser intensity noise should be well pronounced or even
purposely increased, rather than suppressed. The laser beam is not
supposed to be tightly focused on the sample to reduce the number
of spins participating in formation of the signal. At the same
time, the light power density on the sample, in these
measurements, should be high enough to provide the required
optical nonlinearity. As has been shown in \cite{Walser,Kozlov},
under sufficiently high power densities, when the Rabi frequency
becomes comparable with relaxation rates of the excited state, the
intensity noise spectrum of the transmitted light becomes much
more complicated, with its peaks and singularities not connected
in a straightforward way with the Zeeman and Rabi frequencies of
the system.

This technique of noise spectroscopy, as far as we know, was
applied so far only to atomic systems. As the noisy light sources,
in those experiments, were used diode lasers with the
frequency-modulated output emission which was converted into the
intensity-modulated light in the process of its interaction with
narrow spectral features of the sample under study. Meanwhile, the
``active'' noise spectroscopy, in our opinion, is a promising
method of research that may find application in experimental
studies of solid-state (including semiconductor) systems highly
important for the up-to-date photonics and information science.
The experimental setups, with the laser sources intentionally
modulated in intensity or in polarization in a broad frequency
range may be useful in the cases when the nonperturbative nature
of the SNS is not of primary importance. It should be noted that
this method of nonlinear optics may provide information
inaccessible to the conventional SNS and lying far beyond the
bounds of the field of magnetic resonance and spin dynamics.

\subsection{Starting with semiconductors}

Atomic gases seemed to be highly favorable objects of the SNS due
to their intense and narrow optical resonances providing strong
peaks of FR in the vicinity of the lines (the so-called
Macaluso-Corbino effect). Semiconductors and other solid-state
 systems with their broader optical spectra did not look so
promising. Still, in 2005, Oestreich et al. \cite{Oestreich:2005}
have managed to successfully apply this technique to a solid
semiconductor system.

The measurements were performed on a thick $n$-doped GaAs wafer at
10 K with the wavelength of the probe laser $\sim 10$ nm below the
GaAs bandgap. The level of Si-doping ($\sim 1.8\times 10^{16}$ cm$^{-3}$) was chosen corresponding to the longest spin dephasing time measured using the femtosecond time-resolved Faraday rotation technique \cite{Aw}. The FR noise of the sample placed into a transverse
magnetic field was detected with a broadband balanced
photoreceiver and conventional sweeping spectrum analyzer in the
frequency range of 200 - 400 MHz.  Behavior of the discovered
peaks well correlated with the known magnetic-resonance and
relaxation properties of the donor-bound electrons in $n$-GaAs.

This work was, at that moment, more of fundamental than practical
significance since the accumulation time needed to reliably detect
the noise signal lied in the range of several hours. Still, in
these experiments, the authors have successfully used, for the
first time, the SNS technique for measuring the electron Lande
$g$-factor and electron spin relaxation time in a
semiconductor ($n$-GaAs) and have demonstrated applicability of SNS to
semiconductor systems, thus laying the foundation for the semiconductor SNS (see the review \cite{Muller:2010b}).

\section{Technical advancements in spin noise spectroscopy}

The ``academic'' period of the SNS was completed when it became
possible to perform these measurements for much shorter times in a
wider frequency range on system of greater practical importance.

\subsection{Advent of the real-time fast Fourier transform spectrum analyzer}

A real breakthrough in this field of research was made when the
sweeping spectrum analyzer, in the system of data acquisition, has
been replaced by the one with the real-time fast Fourier transform
(FFT) processing system with a wide frequency range
\cite{Romer,Crooker:2010}. The conventional sweeping spectrum
analyzer, as is known, measures the signal only in a narrow
frequency interval at a time and, therefore, constantly disregards
most information containing in the time-dependent signal. The FFT
spectrum analyzer, on the contrary, is capable of using the whole
bulk of incoming information. It digitizes the signal in the whole
bandwidth of the system (with a sampling rate of around $10^9$
s$^{-1}$), performs FFT of the signal {\it in real time}, and
accumulates the spectrum thus obtained. In other words, it
accumulates signals in all frequency channels simultaneously,
rather than only in a single channel. As a result, the
accumulation time needed to achieve the same signal-to-noise ratio
has decreased by a few orders of magnitude. The process of accumulation of the signal
is schematically illustrated by Fig. 11.

\begin{figure}
\begin{center}
\includegraphics[width=10cm]{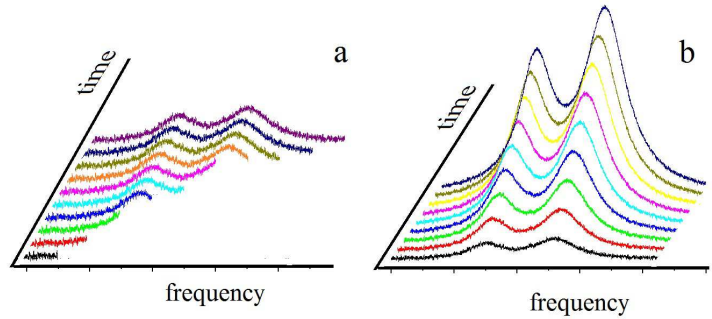}
\caption{The picture shows schematically how spectrum of a broadband signal is accumulated in the scanning (a)
and real-time FFT spectral analyzer (b). In the first case, the detection
system successively passes through each frequency channel ignoring, at that moment, all other channels. In the second
case, signals of all frequency channels are accumulated simultaneously. } \label{fig8}
\end{center}
\end{figure}

This technical advancement has turned the SNS into a real
practical tool of magnetic spectroscopy and made it suitable, in
particular, for studying spin dynamics of low-dimensional
semiconductor systems (quantum wells, quantum dots, quantum wires)
highly important for the present-day applications in photonics and
optoelectronics.

\subsection{Expanding the detection bandwidth}

Another interesting idea was proposed by M\"uller et al.
\cite{Muller:2010a} to overcome bandwidth limitations of the
optical detectors. For that purpose, it was proposed to use a
pulsed laser with a high repetition rate (e.g., a mode-locked
laser) instead of a cw laser, as a source of the probe light. In
this case, the intensity spectrum of the light comprises a comb of
discrete lines spaced by the pulse repetition rate $f_0$, and it
becomes possible to observe, in the spectrum of the detected FR
noise signal, not only the peak at the frequency of magnetic
resonance $f_R$, but also the peaks at the frequencies shifted
from $f_R$ towards lower frequency by multiples of the repetition
rate $f_0$ (provided that $f_R > f_0$). As a result, by mixing the
resonance signal with the nearest peak of the probe beam intensity
spectrum, the frequency of the spin noise resonance can be
transferred to the frequency range $[0, f_0/2]$. This experimental
approach allowed the authors to detect spin noise at Larmor
frequencies up to 16 GHz. Under these conditions, the total
bandwidth of the detection system (including spectrum analyzer)
may not exceed the repetition rate $f_0$. At the same time, the
measured spin dephasing rates are limited, in this technique, by
approximately half the repetition rate. As noted in
\cite{Muller:2010b}, the sensitivity of this technique is not
reduced as compared to the conventional SNS.

\begin{figure}
\begin{center}
\includegraphics[width=10cm]{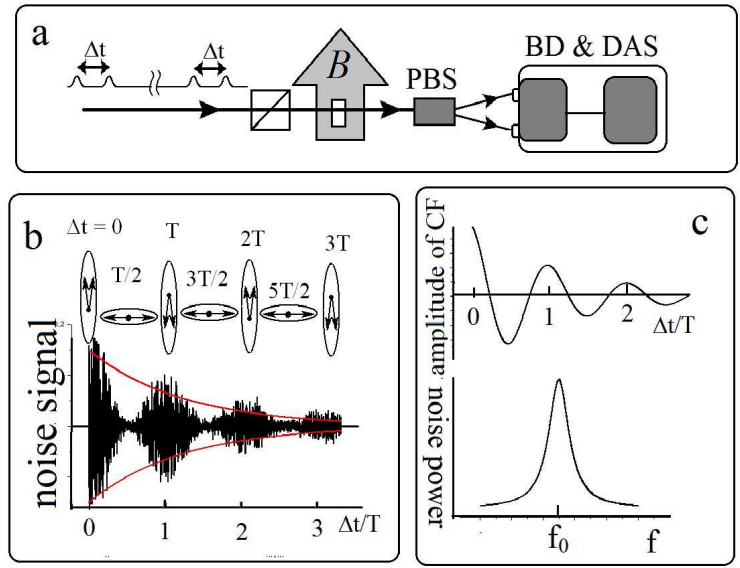}
\caption{Ultrafast spin noise spectroscopy. (a)-– Schematic of the experimental setup. The sample is probed by a train of pairs of ultrashort pulses with variable delay $\Delta t$. (b)- Dependence of the noise signal on the time delay. Vectorial diagrams show summation of contributions of the two pulses to the FR signal for different ratios of the time delay $\Delta t$ and the Larmor precession period $T$ . (c) – The resulted dependence of the noise signal on the time delay  $\Delta t$, corresponding to resonant transient of the system or its autocorrelation function, and the spin resonance spectrum obtained by Fourier transform of the autocorrelation function.  } \label{fig8}
\end{center}
\end{figure}

One more remarkable approach to the problem of bandwidth of the
FR-based SNS was proposed in \cite{Starosielec}. The idea of the
approach was to take advantage of an extremely broad spectrum of
the femtosecond (or picosecond) light pulses emitted by the
mode-locked Ti-sapphire laser and to measure correlation
characteristics of the FR noise with a very high temporal
resolution inaccessible to conventional photodetectors. For that
purpose, the sample under study was suggested to be probed by a train
of  pairs of ultrashort pulses with the time interval between them
variable within the range comparable with the period of Larmor
precession of the system (Fig.12,a). In spite of the fact that the
two closely spaced pulses are not resolved by the detection
system, the detected FR noise will depend on the time interval
between them (provided that there is some distinguished
oscillation frequency (and, therefore, a distinguished characteristic time)
in the FR noise, as shown in Fig. 12,b).  In
particular, when  the time interval between the pulses is equal to
integer number of the oscillation periods, then contributions of
this oscillating process to both pulses will have the same sign,
and total FR of this pair of pulses will fluctuate . If, however,
the time interval between the pulses equals odd number of
half-periods of the oscillation, then contributions of these
oscillations to the two pulses will compensate for each other, so
that total contribution of this process to FR of the pair of
pulses will vanish. Thus, by scanning the time delay between the
two pulses of the probe beam and detecting the FR noise power in
the transmitted light, we will observe an oscillatory transient of
the process directly corresponding to its correlation function,
whose Fourier transform provides the autocorresponding spin-noise
spectrum (Fig. 12,c).  It is important that the two pulses are not
supposed to be resolved by the detection system.

The efficiency of this ultrafast SNS has been
recently successfully demonstrated experimentally on a heavily
$n$-doped bulk GaAs \cite{Berski}. The train of picosecond pulses
was produced, in that experiment, by two synchronized lasers with a
repetition rate of 80 MHz.

This technique, that implies measuring autocorrelation function
instead of its Fourier image, should allow one to directly detect
decay of the transverse magnetization in time domain and to extend
the accessible frequency range up to several THz.

The above proposals demonstrating possibility of considerable
extension of the SNS bandwidth give promise that this technique
will find application in the EPR spectroscopy of standard
microwave ranges for nonperturbative investigations of transparent
paramagnets.

\subsection{ Cavity-enhanced spin noise spectroscopy}

In spite of remarkable advances in the data acquisition technique,
the problem of polarimetric sensitivity retains its significance
in the FR-based SNS, and all suggestions that can help to improve
this characteristic remain to be of great interest.

It is well known that Faraday rotation can be strongly enhanced
with the aid of a Fabry-Perot cavities
\cite{Rosenberg,Ling,Kavokin,Salis,Li,Dong,Giri}, due to multiple
passes of the light through the magneto-optical medium. Such an
approach is especially popular nowadays for studying dynamics of
spin states of low-dimensional semiconductor structures in
high-finesse microcavities, where the observed FR can be increased
by a few orders of magnitude. In these studies, it is usually
tacitly assumed that the measured times are much longer than
intrinsic times of the cavity (the cavity photon lifetime and
intermode beat period). This is really the case for microcavities,
when the photon round-trip time over the cavity is equal or just a
few times longer than the oscillation period  of the light wave.
When, however, this is not the case, and the FR oscillation
frequency becomes comparable with or higher than that of the
intermode beats, the response of the cavity becomes more
complicated.

In \cite{Zap:2011}, it was shown that polarization of a
monochromatic light resonant to a longitudinal mode of a
Fabry-Perot cavity, may be highly sensitive to modulation of the
intracavity anisotropy at frequencies multiple of spacing between
its longitudinal modes. This effect can be qualitatively
understood in terms of the light traveling over the cavity back
and forth and has much in common with the known effect of
mode-locking in lasers \cite{Lamb}.

Indeed, intuitively it seems evident that enhancement of the
cavity's response to a weak oscillating optical anisotropy of the
intracavity element will be observed when the light moving back
and forth inside the cavity finds this element with the same phase
of the oscillation, so that the new contribution to the light
polarization state is added to those already accumulated. This
reasoning, though formally incorrect (being inapplicable to a
monochromatic wave), proves to be useful for qualitative
understanding of the resonant FR enhancement effect.

This effect can be also interpreted in terms of spectral
transformation of the light passing through a polarization
modulator: the resonant enhancement of the FR occurs when the
modulation frequency coincides with intermode spacing of the
Fabry-Perot cavity, and the sidebands of the modulated laser light
hit adjacent modes of the resonator.

Figure 13 shows frequency dependence of the polarization modulation
gain factor ($\Gamma$) for two different positions of the
anisotropic element inside the cavity (for more detail, see
\cite{Zap:2011}). For the sample placed in the middle of the
cavity (a), the light hits the sample twice per a round-trip, and,
correspondingly, the effect of enhancement can be observed at
frequencies that are by a factor of two lower than those for the
sample placed at the edge of the cavity (b).

\begin{figure}
\begin{center}
\includegraphics[width=8cm]{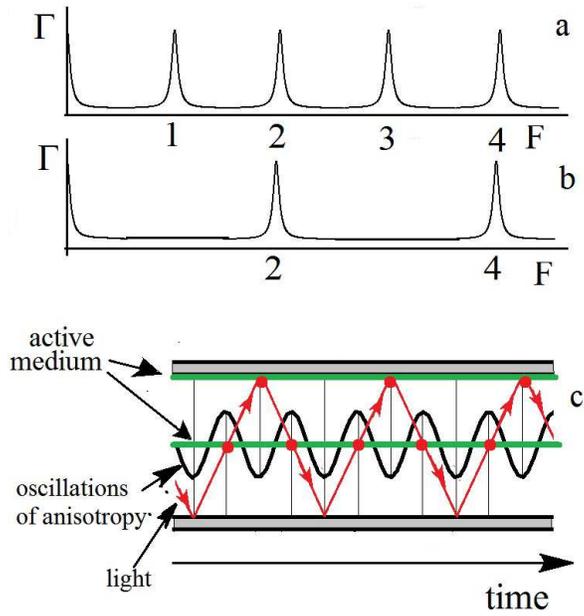}
\caption{
Frequency dependence of the polarization signal gain
factor ($\Gamma$) in a Fabry-Perot cavity for the sample placed at
the edge of the cavity (a) and in the middle of it (b). The
frequency $F$ is given in the units of intermode spacing
$f_0 = c/2L$. Figure (c) illustrates synchronism of the light traveling
over the cavity with oscillations of anisotropy of the intracavity sample
(for $F = 1$). The layers in the center of the cavity and near its
edge (colored green) depict the medium with oscillating anisotropy,
black sinusoid is the time variation of the anisotropy, and red lines
show propagation of the light beam.} \label{fig9}
\end{center}
\end{figure}

Amplification, in this scheme, is realized in the optical (rather
than electronic) channel, and the corresponding gain factor proves
to be approximately equal to the cavity Q-factor ($\sim 10^2 -
10^3$). In other words, the Fabry-Perot cavity can be used as a
{\it selective optical amplifier} of a polarization signal
produced by intracavity element at the frequency of intermode
spacing ($f_0 = c/2L$, $L$ is the cavity length and $c$ is the
speed of light) or its multiples ($f = nf_0$, $n$ is the integer).
This effect can evidently be helpful in SNS for amplification of
the polarization modulation arising in the probe light beam due to
the random spin precession of the spin-system.

It is noteworthy that the increase in the FR signal in this method
is achieved, exactly as in the case of high-extinction
polarimetric measurements, at the expense of strongly increasing
power density on the sample, which should be taken into account in
the experiments. .

One more curious possibility of application of this effect is
related to prospects of creating an all-optical spin-noise
spectrometer \cite{Zap:2011}. The idea of this proposal combines
the effect of amplification of the polarization signal with its
optical quadratic detection. Let the studied transparent
paramagnetic sample be placed inside an optical cavity (Fig.14).
A monochromatic linearly polarized laser beam incident upon the
cavity coincides in frequency with one of its longitudinal modes
and thus passes through the cavity with no loss. At the exit of
the cavity, we
\begin{figure}
\begin{center}
\includegraphics[width=9cm]{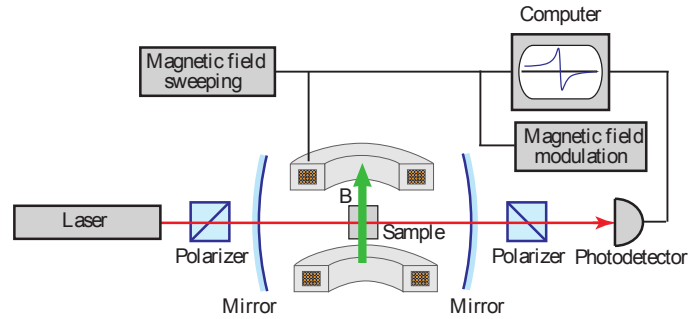}
\caption{ Schematic of the all-optical spin noise spectrometer:
(1) laser, (2) polarizers, (3) mirrors of the optical cavity, (4)
magnetic coils, (5) sample, (6) photodetector, (7) magnetic field
sweeping unit, (8) magnetic field modulation unit, and (9)
detection system. } \label{fig10}
\end{center}
\end{figure}
place a linear polarizer in a crossed position, so that no light
passes through the system. Now, we start sweeping the transverse
magnetic field applied to the sample. At the moment when the
Larmor precession frequency of the spins becomes equal to (or
multiple of) the double intermode frequency of the cavity ($2f_0 =
c/L$), oscillations of the FR appear to be strongly enhanced by
the cavity, and the light intensity at the exit (after the output
polarizer) becomes nonzero.

A unique property of this system is
that it allows one to detect magnetic resonance in the FR noise
spectrum at any frequency without any broadband electronics and
broadband photodetectors - all the needed information is contained
in the dc optical signal. What is observed here is just a result
of resonant coupling of the spin-system with the Fabry-Perot
cavity.

It is clear that such a design of the all-optical spin
spectrometer imposes heavy demands both on Q-value of the cavity
and on extinction ratio of the polarization system. Still, it
seems feasible and may be useful, in particular, for applied
purposes, as a basic system for magnetometers of new generation.

\section{Optical spectroscopy of spin noise}

A considerable interest has been currently attracted to
particularities of spin-noise spectroscopy under conditions of
resonant or near-resonant optical probing of the spin-system
(\cite{Crooker:2010,Chalupczak,Horn}). It is clear that, generally
speaking, the SNS, in this case, looses its important property of
being perturbation-free, when the probe laser beam induces real
optical transitions. At the same time, the degree of perturbation
of the spin-system by the probe beam depends on how large is the
optical excitation rate compared with the dephasing rate of the
spin-system, and, in most cases, the light power density can be
decreased to the level when optical perturbation of the system may
be neglected (see, e.g., \cite{Huang}). Under these conditions,
the SNS acquires additional remarkable properties
\cite{Zapasskii:2012}.

Interesting possibilities  of SNS stems from the question about
mutual correlations of the FR noise at different wavelengths of
the probe beam. In \cite{Zapasskii:2012}, it was pointed out that
these correlations depend on whether the appropriate fluctuations
are contributed by the same spin ensemble (ensemble of identical
quantum systems) or not. It is important
that this fact can be revealed not only in a straightforward way
in the cross-correlation spectral characteristics of the FR noise
(which may be thought of),
but also, much easier, in optical spectra of the FR noise power,
which, generally, appear to be related to the conventional FR spectra in a
nontrivial way.

The use of the probe beam wavelength as a tunable parameter of the
standard spin noise spectroscopy allows one to look at this
technique as at a sort of {\it optical spectroscopy}. Such an
approach makes it possible, in certain cases, to detect the
structure of optical transitions hidden in the linear optical or
magneto-optical spectra and, thus, opens new possibilities of the
SNS technique.

Let us consider dependence of the {\it integrated FR-noise} (spin
noise) power on the probe beam frequency. As was already
mentioned, the FR-based SNS exploits ``paramagnetic'' part of the
Faraday rotation to monitor magnetization of the spin-system. It
seems evident that, to make conversion of the spin-system
magnetization to the FR most efficient, one has to select the
wavelength of the probe light in the region of the greatest FR
(greatest Verdet constant), and, vice versa, the FR noise cannot
be detected at the wavelengths where the FR proper turns into
zero. It was shown, however, that this is not the case.

Magnetization of a macroscopic spin-system is created by an
ensemble of individual spins whose optical spectra may be either
identical or different. As a result, the magnetization noise of
the spin-system may be transformed into that of the FR in
different ways.

If optical spectrum of a paramagnet comprises several spectral
features, then the FR angle $\varphi$ at each frequency $\omega$
will be given by the sum of partial contributions of all these
features:
\begin{equation}
\varphi(\omega) = \sum \varphi_i(\omega)
\end{equation}
When all the spins of the system are identical, and optical
spectrum of the whole ensemble just reproduces the one
corresponding to an individual spin-system, then fluctuations of
the FR will be evidently {\it coherent} or {\it correlated} over
the whole spectrum, and spectral dependence of the mean-square
fluctuation of the FR $\langle\delta\varphi^2(\omega)\rangle$ will
be proportional to the square of the total FR (square of sum of
partial contributions):
\begin{equation}
\langle\delta\varphi^2(\omega)\rangle \sim
\langle(\sum\varphi_i(\omega))^2\rangle
\end{equation}
When, however, the paramagnet under study contains several spin
subsystems with different optical spectra, their contributions of
these subsystems to the FR noise will fluctuate {\it
independently}, in uncorrelated way, and optical spectrum of the
FR noise will be described by the {\it sum of squares} of the
individual contributions rather than by their sum squared:
\begin{equation}
\langle\delta\varphi^2(\omega)\rangle \sim \sum
\langle\varphi_i(\omega)^2 \rangle
\end{equation}
As a result, optical spectra of spin noise in these two cases, due
to different interference of partial contributions in the region
of their overlap, may be essentially different. Figure 15
illustrates this difference for two closely spaced absorption
lines associated either with the same spin-system or with
different (independent) ones characterized by different optical
spectra. Spectral dependence of the FR is taken here in the form
of a dispersion-like curve characteristic of  the ``paramagnetic''
part of the FR for the band with the width substantially exceeding
magnetic splitting of the optical transition. One can see that the
distinction between these two spectra (Fig. 15, c,d) is most
\begin{figure}
\begin{center}
\includegraphics[width=9cm]{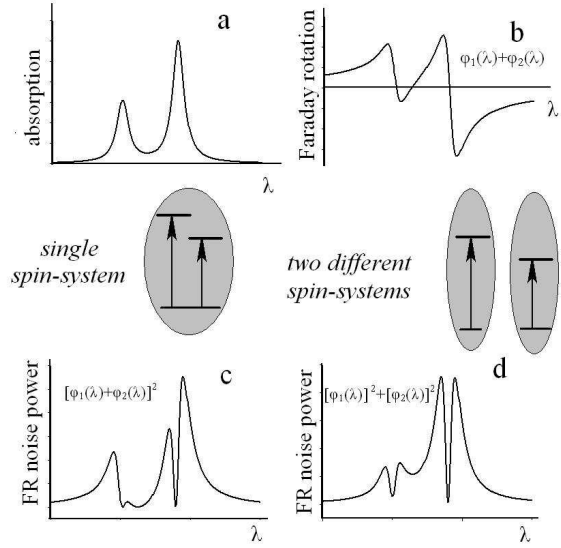}
\caption{ Optical spectra of absorption (a), Faraday rotation (b),
and FR noise power (c and d) of a hypothetical paramagnet with two
closely spaced optical transitions. Spectra c and d correspond to
the cases when optical transitions are associated with the same
spin-system or with two different spin-systems, respectively. }
\label{fig11}
\end{center}
\end{figure}
pronounced in the region between the lines, where contributions of  two optical transitions either compensate for each other (for a
single spin-system, Fig. 15, c) or sum up (in statistical sense,
for two different spin-systems, Fig. 15, d).

One more manifestation of correlation characteristics of spin noise
in the optical spectrum, as noted in \cite{Zapasskii:2012}, is that the optical spectroscopy of spin noise may show, in some cases, spectral resolution higher than the conventional optical or magneto-optical spectroscopy and, thus, may be
helpful in resolving hidden structure of optical spectra.

This effect is revealed in a highly spectacular form in optical
spectra with inhomogeneous broadening \cite{Zapasskii:2012}. The
absorption and FR spectra in the vicinity of an isolated
absorption band (Fig. 16, a,b) are known to be the same regardless
of whether the band is broadened homogeneously or inhomogeneously.
Linear magneto-optics (as well as linear optics in general) cannot
distinguish between these two cases. The results of calculations
presented in \cite{Zapasskii:2012} show that the optical spectra
of spin noise in these two cases are drastically different. (Fig.
16, c,d)

For the homogeneously broadened band, the FR noise, which is
proportional to the Verdet constant squared, vanishes at the
center of the band (where the Faraday rotation proper turns into
zero), whereas for the band with a strong inhomogeneous
broadening, this central dip disappears, and the FR noise proves
to be the
\begin{figure}
\begin{center}
\includegraphics[width=9cm]{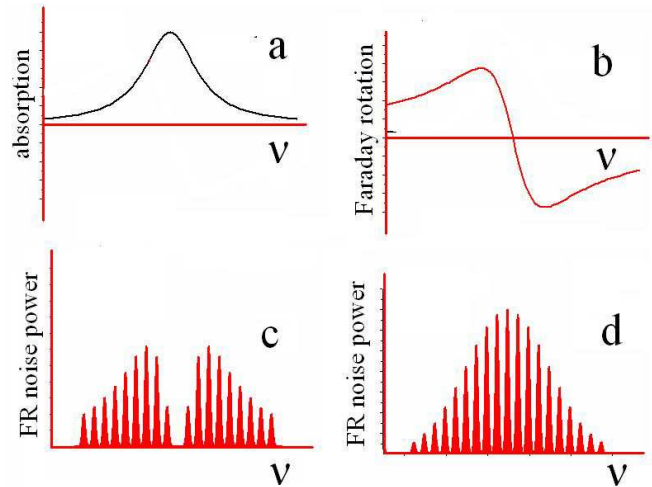}
\caption{ Typical spectra of optical absorption (a), Faraday
rotation (b), and Faraday rotation noise for the cases of
homogeneously (c) and inhomogeneously (d) broadened band. The two
lower plots show schematically how the spin noise spectra (red
peaks) vary with the optical frequency of the probe beam ($\nu$).}
\label{fig12}
\end{center}
\end{figure}
greatest at the band center. Qualitatively, it can be easily
understood. The Faraday rotation at the center of the
inhomogeneously broadened band vanishes because positive and
negative contributions of the higher- and lower-lying spectral
components compensate for each other, whereas their fluctuations,
being uncorrelated, are summed up statistically and attain the
greatest value at the point where the density of spectral
components is the greatest, i.e., at the center of the band.

When the inhomogeneous broadening is comparable with homogeneous,
the dip in the optical spectrum of the FR noise power acquires
some intermediate depth, which can be used to estimate the ratio
of these two contributions to the line broadening.

It was also established in \cite{Zapasskii:2012} that magnitude of
the spin noise considerably increasers with increasing ratio of
the inhomogeneous bandwidth to homogeneous (approximately in
direct proportion with it). It results from the fact that, in the
case of strongly inhomogeneously broadened band, the main
contribution to the FR noise of the probe beam is made by the
spectral components lying nearby the laser wavelength (within the
range of homogeneous width). This relatively small number of
narrow spectral components with relatively large partial
contribution of each of them may provide strong enhancement of the
spin noise power for inhomogeneously broadened bands. This fact,
on the one hand, makes easier SNS experiments with inhomogeneously
broadened systems (like quantum dots) and, on the other, can be
used for measuring homogeneous linewidth of optical transitions.

The above properties of the optical spin-noise (OSN) spectroscopy
have been confirmed experimentally in \cite{Zapasskii:2012}. The
OSN spectra of the homogeneously broadened D1 line of potassium atoms and inhomogeneously broadened band of the InGaAs/GaAs
quantum dots were strongly different in accordance with the above
results. The effect of enhancement of the spin noise power in inhomogeneously broadened systems was also confirmed experimentally.

Thus, the optical spectroscopy of the FR noise may be considered as the other
side of the FR-based SNS that may provide interesting additional
information about spin-system under study.

\section{Unique properties of spin noise spectroscopy}

The SNS, as a method of the ESR spectroscopy, was primarily
intended for providing standard information about g-factors and
relaxation rates of the spin-system. As an optical technique, the
FR-based SNS has much in common with the conventional optical
methods of the ESR detection \cite{Brossel,Aleks:1985,Zap:1987},
with an essential difference that the detected spin precession, in
the SNS, is spontaneous (stochastic) rather than coherently
excited by an external field, and the spin-system is supposed to
remain in the state of thermal equilibrium. What is highly
significant, in our opinion, is that the use of magnetization {\it
noise} in the capacity of signal, in combination with the optical
(laser-assisted) technique, allows one to get additional
information fundamentally inaccessible for conventional methods of
spectroscopy and makes the SNS unique in many respects.

One of the most important features of the FR-based SNS, as was
already mentioned, is related to its nonperturbative character:
the light interacting with the paramagnet in the region of its
transparency produces virtually no real excitation of the system.
This property may be highly important, e.g., in studies of
ultracold atoms \cite{Mihaila:2006b} or semiconductor systems,
when appearance even of small amount of photo-induced charge
carriers may substantially distort dynamics of the spin-system.

Another property of the SNS that essentially distinguishes it from
other (fundamentally perturbative) methods of the
magnetic-resonance spectroscopy is that it does not require
population difference between magnetic sublevels of the
spin-system to detect the resonance. This technique remains
equally efficient at high temperatures and low magnetic fields
(including zero field), when the population difference becomes
negligibly small (see, e. g.,
\cite{Aleks:1981,Crooker:2009,Crooker:2010}.

By detecting spin noise as described above, we, in fact, monitor
random microscopic motion of a macroscopic object confined
spatially by the probe laser beam. This allows us to get
information hardly accessible for the conventional radio- or
linear optical spectroscopy. In particular,
 SNS was used to
identify Fermi-Dirac statistics of a degenerate electron gas in
heavily doped n-GaAs, to distinguish between localized and
delocalized conduction-band electrons, and to detect the effect of
Brownian motion of the electrons in the conduction band on the
spin noise line width. Essential information can be also extracted from
absolute value of the spin noise power which is known to be
directly related to total number of spins probed by the beam \cite{Crooker:2009, Muller:2008,Muller:2010b}.

A widely-known method of nonlinear optics is the so-called
"$Z$-scan technique", which allows one to identify optical
nonlinearity and to measure, in a simple way, coefficients of
nonlinear absorption or refraction of the medium
\cite{Sheik,Patterson}. In all modifications of this
\begin{figure}
\begin{center}
\includegraphics[width=9cm]{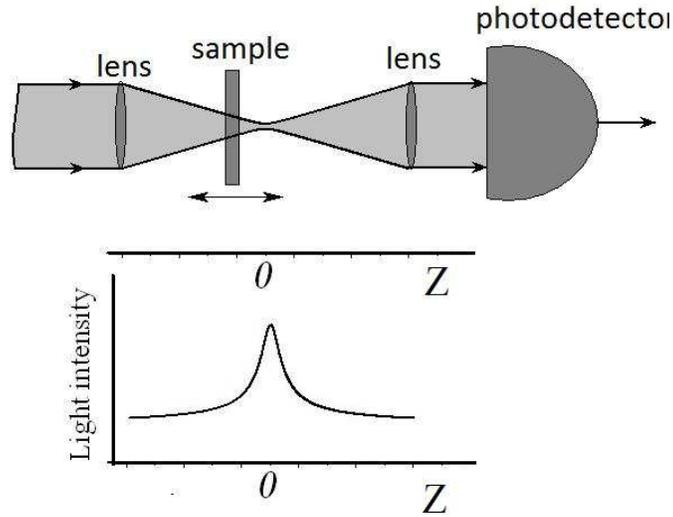}
\caption{ Schematic of the $Z$-scan arrangement (a) and
characteristic dependence of the transmitted light intensity on
position of the sample $Z$ for the sample with optical
nonlinearity.} \label{fig13}
\end{center}
\end{figure}
technique, the sample is drawn through the waist of a tightly
focused laser beam (along the $Z$-axis, Fig. 17), and intensity of
the transmitted light is measured as a function of $Z$-coordinate.
When the sample is optically linear and its optical properties do
not depend on the light power density, the transmitted light
intensity does not show any dependence on $Z$. The presence of
such a dependence with a peculiarity at $Z = 0$ usually serves as
an indicator of nonlinearity of the medium. If we apply this
$Z$-scan technique to the measurements of spin noise, we will
evidently obtain a $Z$-dependent noise signal (as if the medium
were nonlinear), because, when moving the sample through the waist
of the beam, the number of probed spins, controlling magnitude of
the spin noise power, changes. In other words, the tightly focused
laser beam traveling through a bulk paramagnet probes the
spin-system mainly by a small spatial region in the vicinity of
its waist. This situation has evidently much in common with the
case of nonlinear medium when the main response is provided by the
region of the beam with the greatest light power density.

Another curious illustration of this effect may be provided by a simple two-beam experiment, which can be considered as a sort of pump-probe spectroscopy of the light intensity noise (Fig.18). Let a light beam pass through a layer of transparent medium and let us ask the question; Can we find the spot illuminated by this beam on the layer with the aid of another light beam? The usual reply is: Yes, we can do this provided that the first beam (usually called ``pump’’) changes in some way optical properties of the layer in the illuminated spot. In other words, this is possible (and is a usual story) in {\it nonlinear} optics. In the light intensity noise, however, this is also possible, in spite of the fact that the ``probe’’ beam is not supposed to modify, in any way,  properties of the layer.

\begin{figure}
\begin{center}
\includegraphics[width=10cm]{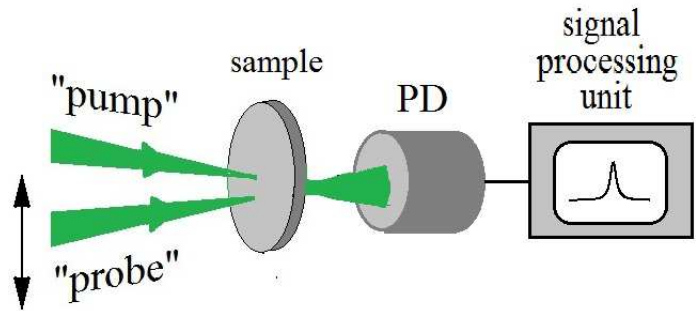}
\caption{Schematic of a two-beam intensity-noise-based experiment demonstrating detection of a spot illuminated by a ``pump'' beam with the other beam (``probe'') scanning over the sample layer. The noise modulating the light beam is supposed to be caused by microscopic dynamics of the illuminated spot. } \label{Forr}
\end{center}
\end{figure}

Indeed, if we detect total intensity noise of the two beams (``pump’’ and ``probe’’) transmitted by the layer (Fig. 18), then the measured signal will depend on whether they pass through the same spot of the layer or not. In the first case, their fluctuations will be correlated, and total noise power will be given by their sum squared, while, in the second case, they will be uncorrelated, and the measured noise power will be given by the sum of their squares. As a result, the spot illuminated by the ``pump’’ can be easily detected by the ``probe’’. We can say, that the illuminate spot is coded in a unique way, and the hey to its code is provided  by the noise of the ``pump ‘’.

It is appropriate to mention here the idea of two-beam spin noise spectroscopy, which was put foeward in \cite{Pershin} and may be promising for studying spatial characteristics of spin systems.

These features of the SNS have been used in \cite{Romer:2009} to
realize a three-dimensional SNS-based tomography. Efficiency of
this technique was demonstrated with a pair of $n$-doped GaAs
wafers, around 350 $\mu$m thick each, probed, as in the
conventional Z-scan technique, by a focused laser beam. The
wavelength of the probe beam was chosen well below the bandgap of
GaAs (849 nm). The measurements were performed with no external
magnetic field, so that the spin noise spectrum was centered at
zero frequency. Due to different doping concentrations of the two
plates, the corresponding spin relaxation times (and spin-noise
spectra) were different, which made it possible to distinguish
them in the SNS experiment. In this proof-of-principle experiment,
the spatial in-depth resolution of 50 $\mu$m was demonstrated. The
method may allow, in authors' opinion, to realize 3D doping
measurements with submicrometer spatial resolution even at low
doping concentrations. This is one of the features of SNS that,
along with its ability to penetrate inside hidden structure of
optical transitions, mentioned above, bring is close to the
methods of nonlinear optics.

\section{Conclusions}
We have briefly outlined development of the SNS for
the last several years and described its main achievements and
potentialities as applied to scientific research. It is curious
that the idea of detecting magnetic resonance in the
Faraday-rotation noise spectrum, that looked initially more like
an academic trick useful primarily for tutorial purposes, gave
birth to a highly efficient and, in many respects, unique
experimental tool. This technique keeps certain properties of
conventional experimental methods like magnetic resonance
spectroscopy, optical and Raman spectroscopy, spectroscopy of
double RF-optical resonance, and Gorter's paramagnetic relaxation
method, but essentially differs from any of them and acquires
thereby qualitatively new features, some of which were considered
in this paper.

Nowadays, as we believe, the FR-based spin noise spectroscopy, as
a new experimental tool, is only at the beginning of its life in
experimental research, and its potentialities are far
from being exhausted. Especially promising looks its application
to studies of micro-samples and nanostructures, bearing in mind
that smallness of the optically probed volume of the sample
(smallness of the number of spins) may be considered, under
certain conditions, as a favorable factor from the viewpoint of
relative magnitude of the signal. An important degree of freedom
of the SNS, which may considerably widen its informative capacity,
is provided by the wavelength of the probe light. Nontrivial shape
of optical spectra of the spin noise power may contain information
about hidden structure of optical transitions and thus may reveal
apparent features of nonlinear optics. Interesting possibilities
may be provided by different modifications of the intentionally
perturbative (resonant) methods of the SNS. These methods,
generally, do not have much to do with polarization measurements
and do not need high polarimetric sensitivity. Their main
advantage is similar to advantage of the FFT spectrum analyzer
compared to the sweeping one: the broadband (rather than
monochromatic) intensity modulation of the acting light
substantially improves sensitivity of the technique. These methods
of the ``active'' noise spectroscopy, however, pertaining more to
nonlinear optics , may provide the information related, to a
greater extent, to the optical (rather than spin) dynamics of the
system.

At present, we have every reason to believe that the novel
technique of SNS will soon find a wide use in studies of
spin-systems and will turn into a standard method of experimental
research. This is a unique case when a pure noise, usually
considered as a  nuisance factor, turns into a basic source of
information in a field of science.

\section*{Acknowledgments}
The author is thankful to Gleb G. Kozlov, Aleksey Kavokin, Evgenii
Aleksandrov, and Dmitri Yakovlev for numerous useful discussions.
The financial support from the Russian Ministry of Education and Science
(contract No. 11.G34.31.0067 with SPbSU and leading scientist A. V. Kavokin) is acknowledged.

%\bibliographystyle{apsrmp4-1}
%\bibliography{nc,cds,sw,connes}
%\bibliography{rmp-sample-zap2}

\end{document}